\documentclass{article}
\begin{document}
\title{\bf Non-Locality and Classical Communication of the Hidden Variable
Theories}
\author{A. Fahmi \footnote{ fahmi@theory.ipm.ac.ir}
\vspace{7mm}\\
{\small Institute for Studies in Theoretical Physics and
Mathematics (IPM)},{\small P. O. Box 19395-5531, Tehran, Iran.}}

\date{{\small{(\today)}}}
\maketitle

\begin{abstract}
In all local realistic theories worked out till now, locality is
considered as a basic assumption. Most people in the field
consider the inconsistency between local realistic theories and
quantum mechanics to be a result of non-local nature of quantum
mechanics. In this paper, we derive Bell's inequality for
particles with instantaneous interactions, and show that the
aforementioned contradiction still exists between quantum
mechanics and non-local hidden variable models. Then, we use this
non-locality to obtain the GHZ theorem. In what follows, we show
that Bacon and Toner's protocol, for the simulation of Bell
correlation, by using local hidden variables augmented by
classical communication, have some inconsistency with quantum
mechanics. Our approach can answer to Brassard questions from
another viewpoint, we show that if we accept that our nature obeys
quantum mechanical laws, then all of quantum mechanic results
cannot be simulated by realistic theories augmented by classical
communication or a single instance use of non-local box.
\end{abstract}


\maketitle
\section{Introduction}
One of the main problems in physics that has attracted physicists'
attention in recent years is locality. This notion has different
meanings, interpretations and applications in different fields of
studies. Physicists consider locality principle as a physical
constraint which should be satisfied by any new theory. Quantum
mechanics (QM) has been challenging this principle for a long
period. Non-locality in QM, however, enters into calculations as a
consequence of the entanglement between some appropriate degrees
of freedom of two separated particles, which makes them to show a
correlated behavior. However, an exact quantitative relation
between non-locality and entanglement has not been known. The
non-local property of QM was first demonstrated by
Einstein-Podolsky-Rosen (EPR) \cite{EPR}, who explicitly suggested
that any physical theory must be both local and realistic. The
manifestation of these conditions then appeared in the so-called
Bell inequality \cite{Bell}, where locality is a crucial
assumptions violated by quantum mechanical predictions. Bell's
inequality has been derived in different ways \cite{CHSH,CH,KS}.
From the so-called Bell's inequalities one can infer Bell's
theorem which states that: " \emph{In a certain experiment all
realistic local theories with hidden variables are incompatible
with quantum mechanics} ". In Bell's theorem, the locality
assumption was involved quantitatively for the first time.

Greenberger \emph{et al.} \cite{GHZ}, Hardy \cite{Hardy} and
Cabello \cite{Cab} have shown that it is possible to demonstrate
the existence of non-locality for the case of more than two
particles without using any inequality. In a recent paper, Barrett
\emph{et al.} \cite{Barre} used the two-side memory loophole, in
which the hidden variables of $n$ pair can depend on the previous
measurement choices and outcomes in both wings of the experiment.
They have shown that the two-side memory loophole allows a
systematic violation of the CHSH inequality. In another model,
Scarani and Gisin \cite{Gisin} considered some superluminal hidden
communication or influences to reproduce the non-local
correlations. Various experiments have been performed to
distinguish between QM and local realistic theories \cite{As}.
They all give strong indications against local realism.

Others' works on this subject can be summarized as follows: The
extension of Bell's inequality and Greenberger, Horne, Zeilinger
(GHZ) theorem to continuous-variables \cite{Wenger}; The Bell-type
inequality that involves all possible settings of the local
measurement apparatus \cite{Zuk}; The extension of local hidden
variable models (LHV) to multiparticle and multilevel systems
\cite{Cab1}; The violation of Bell's inequality beyond Cirelson's
bound \cite{Cab2}.

Bell proved that quantum entanglement enables two space-like
separated parties to exhibit classically impossible correlations.
Even though these correlations are stronger than anything
classically achievable, they cannot be harnessed to make
instantaneous (faster than light) communication possible.

Some people extend Bell's approach, by considering realistic
interpretation of QM and show that an exact simulation of the Bell
correlation (singlet state) is possible by using local hidden
variables augmented by just one bit of classical communication
\cite{Brass,Bacon}. Hence, C. Caves and co-works presented a
model, motivated by the criterion of reality, put forward by EPR
and supplemented by classical communication, which correctly
reproduces the quantum mechanical predictions for measurements of
all products of Pauli operators in an $n$-qubit GHZ state and an
arbitrary graph state \cite{Cav}.

Hence, Popescu and Rohrlich \cite{Cerf} have shown that even
stronger correlations can be defined, under which instantaneous
communication remains impossible. It has been recently shown that
all causal correlations between two parties which respectively one
bit outputs $a$ and $b$ upon receiving one bit inputs $x$ and $y$
can be expressed as convex combinations of local correlations
(i.e., correlations that can be simulated with local random
variables) and non-local correlations of the form $a+b=x\cdot y$
\hspace{.2cm} \emph{mod 2}. It is also shown that a single
instance of the latter elementary non-local correlation suffices
to simulate exactly all possible projective measurements that can
be performed on the singlet state of two qubits, with no
communication needed at all \cite{Cerf}. Recently, G. Brassard and
co-workers raised the question \cite{Brass} that was repeated
again by Bacon and Toner \cite{Bacon}: \emph{Can we find a Bell
inequality with an auxiliary communication violated by a quantum
state, using a set of quantum measurements?} Hence, they consider
this question from another viewpoint, and ask: Why are the
correlations achievable by quantum mechanics not maximal among
those that preserve causality? They give a partial answer to this
question by showing that slightly stronger correlations would
result in a world in which communication complexity becomes
trivial \cite{Brass1}.

In this Paper, we derive the Bell-type inequality \cite{Bell} and
prove a GHZ-type \cite{GHZ} theorem for particles with
instantaneous (non-local) interactions at the hidden-variable
level. Then, \emph{we show that the previous contradiction still
exists between QM and non-local hidden variable models.} In what
follows, we consider model suggested by Toner and Bacon
\cite{Bacon} and proposed again by Cerf and co-workers in the
another way \cite{Cerf}. In the continuation, we would like to
answer some questions posed by G. Brassard \emph{et al.}
\cite{Brass,Brass1} from another viewpoint, and we show that if we
accept that nature obeys quantum mechanical laws, then all of
quantum mechanic all results cannot be simulated by realistic
theories, augmented by classical communication \cite{Bacon} or a
single use of nonlocal box \cite{Cerf}.

\section{Non-Locality and Hidden Variable Theory}

Let us now apply this to the standard EPR-Bell setup in which a
particle source with total vanishing spin $(S=0)$ emits pairs of
oppositely- directed neutral spin-$\frac{1}{2}$ particles in a
spin singlet state. The state vector of the system of two
particles is:
\begin{eqnarray}
|\phi^{-}\rangle=\frac{1}{\sqrt{2}}(|+-\rangle_{12}-|-+\rangle_{12}),\label{BS}
\end{eqnarray}
where $|+-\rangle_{12}$ indicates that the spin of the first
(second) particle in the $z$-basis is $+\frac{1}{2}$
$(-\frac{1}{2})$. These two particles are separated in opposite
directions and when their separation becomes space-like, two
experimenters (Alice and Bob), measure their spin components along
$\hat{a}$ and $\hat{b}$, respectively. If the result of
measurement on the particle $1$ $(2)$ is called $A$ $(B)$, then,
in a local hidden variable (LHV) model, the locality condition is
defined as:
\begin{eqnarray}
A(a,b,\lambda)=A(a,\lambda), \hspace{.5cm}
B(a,b,\lambda)=B(b,\lambda).
\end{eqnarray}
In the above expression the result of the measurement of the
particle $1$  $(2)$ is independent of parameters and results of
the measurement performed on the particle $2$  $(1)$. The existing
contradiction between QM and LHV models, and also the violation of
Bell's inequality in experiments \cite{As} lead us to conclude
that we may doubt one of the initial assumptions of Bell's
inequality, i.e, locality or reality. Of course, there are some
people who believe that other assumptions and loopholes might be
involved, instead of the locality and reality assumptions
\cite{Barre,As1}.

Some generalizations of this theorem are also available, among
which the GHZ theorem \cite{GHZ} is the most famous one. In the
GHZ theorem, the assumption that the correlation functions
calculated from LHV models are equal to those obtained from QM,
encounters a contradiction. However, it should be noted that the
GHZ-type theorems are just for $n$-particle systems in which
$n\geq3$. Typically in a GHZ argument, a source with zero total
spin emits four spin $\frac{1}{2}$ particles such that their total
state vector is:
\begin{eqnarray}
|\psi\rangle=\frac{1}{\sqrt{2}}(|++--\rangle+|--++\rangle)_{1234}.
\end{eqnarray}
After the particles are separated from each other in a space-like
manner, the spins of the particles are measured along the
directions $\hat{a},\hat{b},\hat{c}$ and $\hat{d}$, respectively.
If the measurement results on particles 1 through 4 are denoted by
$A, B, C,$ and $ D$ respectively, then, the locality condition in
GHZ theorem will be defined as:
\begin{eqnarray}\label{2}
A(a,b,c,d,\lambda)=A(a,\lambda),
\end{eqnarray}
Similar relations hold for $B, C$ and $D$. Similar to the case of
Bell's theorem, the results of different measurements are
independent of each other, and also independent of other
parameters. Although much effort has been done to understand the
role of locality and reality assumptions, due to the existence of
various definitions, a conclusive source of the contradiction is
hard to explore. Also, there is no clear-cut and sharp distinction
between these two conditions. If we take Einstein's definition of
reality (in EPR) into account, then it can be seen that it has a
close relationship to locality. In other words, had we not assumed
locality, speaking about realism (as Einstein considered) would
have been irrelevant. In the models in which instantaneous
interactions are present (such as Bohm's model), reality is
essentially considered in the context of non-local influences.

In the following, we shall give up the locality assumption in LHV
models and then prove the contradiction between QM and hidden
variable models. To do this, let us assume that there is a deeper
hidden variable level which is represented by independent
variables $i$ (for particle 1) and $j$ (for particle 2). We denote
the corresponding distribution functions of each variable by
$p(i)$ and $q(j)$, respectively. We define the results
$A(a,b,\lambda)$ and $ B(a,b,\lambda)$ by averaging over the
deeper-level variables $i$ and $j$ :
\begin{eqnarray}\label{5}
&A(a,b,\lambda)&= \int p(i)di f_{A}(a, \lambda, i)g_{A}(b,
\lambda, i)\nonumber\\&B(a,b,\lambda)&=\int q(j)dj f_{B}(a,
\lambda, j)g_{B}(b, \lambda, j)
\end{eqnarray}
where
\begin{eqnarray}
&-1\leq f_{A}(a, \lambda, i),g_{A}(b, \lambda, i), f_{B}(a,
\lambda, j), g_{B}(b, \lambda, j)\leq 1 \nonumber\\
& -1\leq
A(a,b,\lambda), B(a,b,\lambda)\leq1 \nonumber\\
&\int p(i)di=\int q(j)dj =1 \hspace{.5cm} p(i) \geq0 \hspace{.5cm}
q(j) \geq0
\end{eqnarray}
This type of generalization is not special. Actually, for any
realistic variable we can make such an assumption. Also, these
variables could not be considered just a mathematical
generalization but as something that could convey some
physical degrees of freedom \cite{Bohm1}.  \\
Considering the above relation, one can show that the following
relation is satisfied:
\begin{eqnarray}
&|f_{A}(a, \lambda, i)g_{A}(b, \lambda, i)f_{B}(a, \lambda,
j)g_{B}(b, \lambda, j) - f_{A}(a, \lambda, i)g_{A}(b',
\lambda, i)f_{B}(a, \lambda, j)g_{B}(b', \lambda, j) | \nonumber\\
&+|f_{A}(a', \lambda, i)g_{A}(b', \lambda, i)f_{B}(a', \lambda,
j)g_{B}(b', \lambda, j)+ f_{A}(a', \lambda, i)g_{A}(b, \lambda,
i)f_{B}(a', \lambda, j)g_{B}(b, \lambda, j) |\leq 2.
\end{eqnarray}
By averaging on $i$ and $j$ variables, we get:
\begin{eqnarray*}
&|A(a, b, \lambda)B(a, b, \lambda)- A(a, b', \lambda)B(a, b',
\lambda)| + |A(a', b', \lambda)B(a', b', \lambda)+ A(a', b,
\lambda)B(a', b, \lambda)| \leq 2
\end{eqnarray*}
This equation tells us that every realistic model (generally as a
non-local one) must satisfy the above inequality. Otherwise, we
encounter an intrinsic inconsistency in the realistic model. Using
this relation and averaging this time on $\lambda$, which has the
distribution function $\rho(\lambda)$, we derive Bell's
inequality.\\
 Now, we can extend the above approach to
derive GHZ's theorem. For the hidden variables in this theorem,
the extension is as follows:
\begin{eqnarray}\label{4}
 A(a,b,c,d,\lambda)
 =\int p(i)di f_{A}(a, \lambda,
i)g_{A}(b, \lambda, i)h_{A}(c, \lambda, i)k_{A}(d, \lambda, i)
\end{eqnarray}
Similar relation can be defined for $B, C$ and $D$, which are in
general different from the local form of eq. (\ref{2}). Thus, we
have:
\begin{eqnarray*}
-1 \leq f_{t}(a, \lambda, i), g_{t}(b, \lambda, i), h_{t}(c,
\lambda, i), k_{t}(d, \lambda, i)\leq 1
 \hskip 2cm t=A,B,C,D
\end{eqnarray*}
After some simple algebra (similar to the case of the usual GHZ
arguments) the contradiction between GHZ theorem and QM can be
obtained. Likewise, This type of generalization to nonlocal cases
is not restricted to eq. (\ref{5}). We can generalize the above
approach to arbitrary numbers of deeper hidden-variable levels to
define arbitrary realistic hidden variables theories. We shall
have very general cases, such as:

\begin{eqnarray}\label{9}
&A(a,b,\lambda)&=\int p(i)di f_{A}(a, b, \lambda, i)g_{A}(a, b,
\lambda, i)\nonumber\\&B(a,b,\lambda)&=\int q(j)dj f_{B}(a, b,
\lambda, j)g_{B}(a, b, \lambda, j)\nonumber\\
&f_{s}(a,b,\lambda,i)&=\int p'(l)dl f'_{A}(a, \lambda,
i,l)g'_{A}(b, \lambda, i,l)\nonumber\\&g_{s}(a,b,\lambda,i)&=\int
q'(k)dk f'_{B}(a, \lambda, j,k)g'_{B}(b, \lambda, j,k)
\hspace{1cm} s=A,B
\end{eqnarray}
With arbitrary distribution functions for each variable
represented by $p'(l)$ and $q'(k)$, and using the above nonlocal
form, we can derive Bell's inequality again. This generalization
can be extended to infinite hidden variable levels \cite{Ref}.

There is suggestion by \cite{Hess} that if one takes Bell's
probability densities $\rho(\lambda)$ to be dependent on the
parameters $\lambda^{1}_{a},\lambda^{2}_{b}$ for Alice and Bob
stations respectively, then, one cannot derive Bell's inequality.
We claim that we take $\rho(\lambda)$ to be the form:
\begin{eqnarray}
\rho(\lambda,\lambda^{1}_{a},\lambda^{2}_{b})=\int\sigma(\lambda,\lambda^{1}_{a},\omega)\tau(\lambda,\lambda^{2}_{b},\omega)\upsilon(\omega)d\omega
\end{eqnarray}
then, one can deduce Bell's inequality.

In the next section, we consider another approach, our approach is
nearly similar to the GHZ theorem. We use Toner and Bacon
arguments for the simulation of QM and show that although their
arguments simulate quantum mechanical correlation function
exactly, their approach cannot simulate all of QM results. This
new approach indicates that our approach is not restricted to
equations (\ref{5}) or (\ref{9}) and that deeper concepts are
involved.

\section{Bacon and Toner Simulation of Quantum Correlation function}
Recently, Bacon and Toner \cite{Bacon} calculated classical
resources required to simulate quantum correlations. They have
shown that for the exact simulation of an entangled Bell pair
state, we only need to use local hidden variables, augmented by
just one bit of classical communication (although in this
protocol, we cannot make any distinction between an instantaneous
effect and classical communication). Hence, Cerf \emph{et al.}
\cite{Cerf}, derived same results with the use of non-local
machine.

In the following, we would like answer a question posed by G.
Brassard \emph{et al.} \cite{Brass} and repeated again by Bacon
and Toner \cite{Bacon}. The question is that by using a set of
quantum measurements, one can find a Bell inequality with an
auxiliary communication violated by a quantum state?

We consider the same protocol that was presented by Bacon and
Toner. In that protocol, Alice and Bob share two random variables
$\hat{\lambda}_{1}$ and $\hat{\lambda}_{2}$ which are real three
dimensional unit vectors. These random variables are chosen
independently and distributed uniformly over the unit sphere
(infinite communication at this stage). Their protocol proceeds as
follows: (1) Alice outputs
$\alpha=-sgn(\hat{a}.\hat{\lambda}_{1})$. (2) Alice sends a single
bit $ c \in \{-1, +1\}$ to Bob where
$c=sgn(\hat{a}.\hat{\lambda}_{1})sgn(\hat{a}.\hat{\lambda}_{2})$.
(3) Bob outputs
$\beta=sgn[\hat{b}.(\hat{\lambda}_{1}+c\hat{\lambda}_{2})]$, where
$sgn$ function is defined by $sgn(x)=+1$ if $x \geq 0$ and
$sgn(x)=-1$ if $x < 0$. The joint expectation value
$\langle\alpha\beta\rangle$ can be calculated using the following
relation:
\begin{eqnarray*}
\langle\alpha\beta\rangle=E\{-sgn(\hat{a}.\hat{\lambda}_{1})\sum_{d=\pm1}\frac{1+cd}{2}\times
sgn[\hat{b}.(\hat{\lambda}_{1}+d\hat{\lambda}_{2})]\}
\end{eqnarray*}
where $E\{x\}=\frac{1}{(4 \pi)^{2}}\int d\hat{\lambda}_{1}\int
d\hat{\lambda}_{2}x$.

Here, we extend Hardy's approach \cite{Hardy1} to non-local cases.
We consider the following statements:
\begin{eqnarray}\label{Bell}
A(\hat{a}_{1}, \hat{\lambda}_{1})B(\hat{b}_{2}, c(\hat{a}_{1}),
\hat{\lambda}_{1}, \hat{\lambda}_{2})=-1, \\ \nonumber
A(\hat{a}_{3}, \hat{\lambda}_{1})B(\hat{b}_{2}, c(\hat{a}_{3}),
\hat{\lambda}_{1}, \hat{\lambda}_{2})=-1, \\ \nonumber
A(\hat{a}_{3}, \hat{\lambda}_{1})B(\hat{b}_{4}, c(\hat{a}_{3}),
\hat{\lambda}_{1}, \hat{\lambda}_{2})=-1, \\ \nonumber
A(\hat{a}_{5}, \hat{\lambda}_{1})B(\hat{b}_{4}, c(\hat{a}_{5}),
\hat{\lambda}_{1}, \hat{\lambda}_{2})=-1,\\\nonumber
.............. \hspace{2 cm}\\\nonumber.............. \hspace{2
cm}\\\nonumber.............. \hspace{2 cm}\\\nonumber
A(\hat{a}_{N-1}, \hat{\lambda}_{1})B(\hat{b}_{N},
c(\hat{a}_{N-1}), \hat{\lambda}_{1}, \hat{\lambda}_{2})=-1,
\\ \nonumber
A(\hat{a}_{1}, \hat{\lambda}_{1})B(\hat{b}_{N}, c(\hat{a}_{1}),
\hat{\lambda}_{1}, \hat{\lambda}_{2})=+1.
\end{eqnarray}
where $N$ is even. Similar to Hardy's arguments, we define the
probability $p^{\pm}(\hat{a}, \hat{b})$ for getting the result of
$A(\hat{a}, \hat{\lambda}_{1})B(\hat{a},
\hat{b},\hat{\lambda}_{1}, \hat{\lambda}_{2})=\pm 1$. Then, the
probabilities for each of statements (\ref{Bell}) being true are,
respectively:
\begin{eqnarray}\label{10}
p^{-}_{1}&=&p^{-}(\hat{a}_{1}, \hat{b}_{2}),\\
\nonumber
p^{-}_{2}&=&p^{-}(\hat{a}_{3}, \hat{b}_{2}),\\
\nonumber
p^{-}_{3}&=&p^{-}(\hat{a}_{3}, \hat{b}_{4}),\\
\nonumber p^{-}_{4}&=&p^{-}(\hat{a}_{5}, \hat{b}_{4}),
\\\nonumber &...........&
\hspace{1 cm}\\\nonumber&...........& \hspace{1
cm}\\\nonumber&...........& \hspace{1 cm}\\\nonumber
p^{-}_{N-1}&=&p^{-}(\hat{a}_{N-1}, \hat{b}_{N}),\\
\nonumber p^{+}_{N}&=&p^{+}(\hat{a}_{1}, \hat{b}_{N}).
\end{eqnarray}
Hardy \cite{Hardy1} defined the probability $P$ to be true for all
the statements (\ref{10}). The probability that one or more of the
statements to be false must be less than or equal to the sum of
the probabilities for each individual statement to be false, {\em
i.e.}:
\begin{eqnarray*}
1-P\leq \sum_{n=1}^{N-1}(1-p_{n}^{-}) + (1- p_{N}^{+})
\end{eqnarray*}
After some simple algebra, he derived the following Bell's
inequality:
\begin{eqnarray}\label{11}
\sum_{n=1}^{N-1}p_{n}^{-} + p_{N}^{+} >N-1
\end{eqnarray}
In the above equation $p_{n}^{-}$ and $ p_{N}^{+}$ represent the
probability that each individual statement (\ref{Bell}) is true.
If we consider the singlet state (\ref{BS}) and calculate the
prediction of quantum mechanics, we get:
\begin{eqnarray*}
p^{\pm}(a, b)=\frac{1}{2}[1\mp\cos(a-b)]
\end{eqnarray*}
Hardy has chosen the angles $a_{1}, b_{2}, a_{3},..., b_{N}$ to be
evenly spread, so that:
\begin{eqnarray}
b_{N}-a_{1}=\phi,\hspace{.7cm} b_{n}-a_{n\pm1}=\mp\frac{\phi}{N-1}
\end{eqnarray}
The maximum violation of the inequality (\ref{11}) is obtained
for:
\begin{eqnarray}
\phi=\frac{(N-1)\pi}{N}
\end{eqnarray}
He has also shown that for $N\longrightarrow\infty$, the
probability that all the statements of (\ref{Bell}) are true tends
to $1$ ($P\longrightarrow1$). In other words, for any measurement
of two-particle systems all the statements (\ref{Bell}) are
correct.

If we insert Alice and Bob outputs in the equation (\ref{Bell}),
all the statements change to:
%\begin{widetext}
\begin{eqnarray}
-sgn(\hat{a}_{2i\mp1}.\hat{\lambda}_{1})[\frac{1-c(\hat{a}_{2i\mp1})}{2}
sgn(\hat{b_{2i}}.(\hat{\lambda}_{1}-\hat{\lambda}_{2}))
%\\\nonumber
+\frac{1+c(\hat{a}_{2i\mp1})}{2}
sgn(\hat{b_{2i}}.(\hat{\lambda}_{1}+\hat{\lambda}_{2}))]&=&-1,\nonumber
\hspace{.7cm}
\\ \nonumber
i=1,...,\frac{N}{2}-1 \\ \nonumber
-sgn(\hat{a}_{N-1}.\hat{\lambda}_{1})[\frac{1-c(\hat{a}_{N-1})}{2}
sgn(\hat{b_{N}}.(\hat{\lambda}_{1}-\hat{\lambda}_{2}))
%\\\nonumber
+\frac{1+c(\hat{a}_{N-1})}{2}
sgn(\hat{b_{N}}.(\hat{\lambda}_{1}+\hat{\lambda}_{2}))]&=&-1,
\\\nonumber
-sgn(\hat{a}_{1}.\hat{\lambda}_{1})[\frac{1-c(\hat{a}_{1})}{2}
sgn(\hat{b_{N}}.(\hat{\lambda}_{1}-\hat{\lambda}_{2}))
%\\\nonumber
+\frac{1+c(\hat{a}_{1})}{2}
sgn(\hat{b_{N}}.(\hat{\lambda}_{1}+\hat{\lambda}_{2}))]&=&+1.
\end{eqnarray}
%\end{widetext}
If we multiply l.h.s. and r.h.s., we obtain:
%\begin{widetext}
\begin{eqnarray}
\{\prod_{i=1}^{\frac{N}{2}-1}\prod_{k=\pm1}[\frac{1-c(\hat{a}_{2i
+k})}{2} sgn(\hat{b_{2i}}.(\hat{\lambda}_{1}-\hat{\lambda}_{2}))
%\\\nonumber
+\frac{1+c(\hat{a}_{2i+k})}{2}
sgn(\hat{b_{2i}}.(\hat{\lambda}_{1}+\hat{\lambda}_{2}))]\}
\\ \nonumber
\times[\frac{1-c(\hat{a}_{N-1})}{2}sgn(\hat{b_{N}}.(\hat{\lambda}_{1}-\hat{\lambda}_{2}))
%\\\nonumber
+\frac{1+c(\hat{a}_{N-1})}{2}sgn(\hat{b_{N}}.(\hat{\lambda}_{1}+\hat{\lambda}_{2}))]
\\ \nonumber
\times[\frac{1-c(\hat{a}_{1})}{2}sgn(\hat{b_{N}}.(\hat{\lambda}_{1}-\hat{\lambda}_{2}))
%\\\nonumber
+\frac{1+c(\hat{a}_{1})}{2}sgn(\hat{b_{N}}.(\hat{\lambda}_{1}+\hat{\lambda}_{2}))]=-1
\end{eqnarray}
%\end{widetext}
If we choose $\hat{\lambda}_{1}=-\hat{\lambda}_{2}$ (noting the
Toner and Bacon protocol, Alice and Bob have unlimited access to
$\hat{\lambda}_{1}$ and $\hat{\lambda}_{2}$), we get
$c(\hat{a})=-1$ (for all statements). Thus, the product of all
these equations must be equal to $+1$ on the l.h.s., and equal to
$-1$ on the r.h.s.. As we said, for $N\longrightarrow\infty$, all
statements of equation (\ref{Bell}) must be true for all
$\hat{\lambda}_{1}, \hat{\lambda}_{2}$. Thus, one gets an
inconsistency between Bacon and Toner's protocol and quantum
mechanics.

Before finishing this section, we would like to explain some
points. Our extension can be considered not only for classical
communication but also for instantaneous effects. Only in the
explanation of protocol, we have a difference between an
instantaneous interaction and a classical communication, but at
the mathematical level, we have no difference. In other words, we
can consider an instantaneous effect that changes Bob's results as
mentioned before. Recently, this non-local approaches was
suggested by Cerf and co-workers, who derived the Bacon and Toner
results with the replacement of classical communications with a
single instance of the non-local machine \cite{Cerf}.

\section{Discussion}

Bell's theorem \cite{Bell} states that any local realistic view of
the world is incompatible with QM, while, this is often
interpreted as demonstrating the existence of non-locality in QM
\cite{Bohm}. There exist some types of models that simulate
quantum correlation function. In these models, quantum correlation
function simulate exactly all possible projective measurements
that can be performed on the singlet state of two qubits, by using
local hidden variables augmented by just one bit of classical
communication or by a single instance of the non-local machine
(without any communication needed at all). The amounts of
classical communication (one instance of nonlocal Popescu-Rohrlich
machine) has been considered as the amount of  the non-locality
inherent in quantum mechanics at entanglement in the singlet
state. Although, these theories explain some part of QM, but they
could not proposes a complete description of QM. For example, in
the non-local machine, we have not all of QM properties. It has
been shown that entanglement swapping are not simulated by
non-local machine \cite{Short} and quantum multiparties
correlation arising from measurements on a cluster state cannot be
simulated with $n$ non-local boxes, for any $n$ \cite{BP}.

By following these works, G. Brassard and co-workers have
propounded an important questions \cite{Brass}: have a Bell
inequality been fund with auxiliary communication that is violated
by a quantum state and a set of quantum measurements? Hence, they
given responses to two other important questions  \cite{Brass1}:
$(1)$ Considering that perfect nonlocal box's would not violate
causality, why do the laws of quantum mechanics only allow us to
implement nonlocal box's better than anything classically
possible, yet not perfectly? $(2)$ Why do they provide us with an
approximation of nonlocal box's that succeed with probability
$\wp=85.4$ rather than something better? They give a partial
answer to this questions by showing that in any world in which
communication complexity is nontrivial, there is a bound on how
much nature can be nonlocal. This bound, which is an improvement
over the previous knowledge that nonlocal boxes could not be
implemented exactly, approaches the actual bound $\wp=85.4$,
imposed by quantum mechanics. The obvious open question is to
close the gap between these probabilities. A proof that nontrivial
communication complexity forbids nonlocal boxes to be approximated
with probability greater than $\wp$, would be very interesting, as
it would render Tsirelson's bound \cite{Ts} inevitable, making it
a candidate for a new information-theoretic axiom for quantum
mechanics.

We have responded to G. Brassard and co-workers important
questions \cite{Brass1} in another way. We have considered hidden
variable theories augmented by classical communication
\cite{Bacon} or a single use of nonlocal box \cite{Cerf} and have
shown that these theories cannot simulate QM (not Bell's
correlations) by local hidden variables augmented by classical
communication or by nonlocal effects. We don't claim that Toner
and Bacon protocol is wrong - that protocol simulates Bell's
correlation functions \emph{exactly} - but it isn't enough. In
other words, \emph{it cannot simulate all of QM results without
internal inconsistency.}

%1) We cannot deduce non-locality from Bell's inequality and their
%extension.

%Similar to BQM which predict QM correlation functions exactly,
%but, in the hidden variable level that must explain exactly what
%is reality or quantum potential, hence, \emph{that must have
%internal consistency}.

In this Paper, we have considered hidden variable theories with
instantaneous (non-local) interactions and have shown that similar
to local hidden variable theories, there exists an incompatibility
between QM and non-local hidden variables theories. Hence, we have
answered Brassard questions from another viewpoint, we have shown
that if we accept that nature obeys quantum mechanical laws, then
all of quantum mechanical results cannot be simulated by realistic
theories augmented by classical communication or a single use of
nonlocal box.

Therefore, it can be concluded that some other alternative view
points might be involved. Some people conclude that the assumption
of the existence of a reality independent of observation may be
irrelevant to physics \cite{Zei1}. On the other hand, some people
still believe that QM is a local theory \cite{Peres}, and some
others consider information the root of the interpretation of QM
\cite{Zei}. Another non-local hidden variable model is Bohmian
quantum mechanics (BQM) \cite{Hol} (although, this theory does not
provide a complete theory for the spin variable). Could one apply
our extension to BQM? Anyway, the above argument indicates that we
must have a deeper understanding of the notions of locality,
reality and entanglement.

{\bf Acknowledgment}: We would like to thank A. Shafiee, M.
Golshani and A. Peres for useful comments and A. T. Rezakhani for
critical reading of the manuscript. (this work supported under
project name: \emph{Gozar Az Khorshid}).

%%%%%%%%%%%%%%%%%%%%%%%%%%%%%%%%%%%%%%%%%%%%%%%%%%%%%%%%%%%%%%%%%%%%%%%%%%%%

%\begin{references}


\begin{thebibliography}{99}
\bibitem{Bohm1} For example in ref \cite{Bohm}, p. 140, the authors have defined some sets of hidden
variables where the result of measurement depends on them (hidden
variables $\mu_{a}$ and $\mu_{b}$ associated with the
corresponding pieces of measuring apparatus of first and second
particles respectively. In addition there will be some sets of
variables $\lambda_{A}$ and $\lambda_{B}$, belonging to the
particle $A$ and $B$ themselves).
\bibitem{Ref} Also, it is worthy to note that one could even go further and
extend the above non-local approach to other relevant theorems
such as CH, Kochen-Specker \cite{CH,KS} and Hardy \cite{Hardy1}
theorems. However, it is an open problem whether one can construct
a hidden variable model, satisfying the above requirements, to
show the consistency of our argument in a more concrete manner.
\end{thebibliography}
\end{document}